\documentclass[fleqn,10pt]{wlscirep}
\usepackage{graphicx}
\usepackage{dcolumn}
\usepackage{bm}
\usepackage{color}
\usepackage{relsize}
\renewcommand{\vec}[1]{\mathbf{#1}}
\usepackage{hyperref}
\newcommand{\scalefac}{1.}

\title{Self-Polarizing Microswimmers in Active Density Waves}

\author[1,*]{Alexander Geiseler}
\author[1,2,3]{Peter H\"anggi}
\author[4,5]{Fabio Marchesoni}
\affil[1]{Institut f\"ur Physik, University of Augsburg, D-86159, Germany}
\affil[2]{Nanosystems Initiative Munich, Schellingstra{\ss}e 4, D-80799 M\"unchen, Germany}
\affil[3]{Department of Physics, National University of Singapore, 117551 Singapore, Republic of Singapore}
\affil[4]{Center for Phononics and Thermal Energy Science, School of Physics Science and Engineering, Tongji University, Shanghai 200092, People's Republic of China}
\affil[5]{Dipartimento di Fisica, Universit\`{a} di Camerino, I-62032 Camerino, Italy}

\affil[*]{alexander.geiseler@physik.uni-augsburg.de}

\begin{abstract}
An artificial microswimmer drifts in response to spatio-temporal modulations of an activating suspension medium. We consider two competing mechanisms capable of influencing its tactic response: angular fluctuations, which help it explore its surroundings and thus diffuse faster toward more active regions, and self-polarization, a mechanism inherent to self-propulsion, which tends to orient the swimmer's velocity parallel or antiparallel to the local activation gradients. We investigate, both numerically and analytically, the combined action of such two mechanisms. By determining their relative magnitude, we characterize the selective transport of artificial microswimmers in inhomogeneous activating media.
\end{abstract}
\begin{document}

\flushbottom
\maketitle
\thispagestyle{empty}

\section*{Introduction}

The tactic movements of self-propelling microorganisms in response to external stimuli are classified based on the type of the activating stimulus and on whether the organism drifts towards (positive) or away from (negative) the stimulus source \cite{Murray}. Tactic cell migration plays a central role in the formation and organization of larger biological organisms
\cite{Berg,Gompper}. Artificial microswimmers are specially designed synthetic microparticles capable to propel themselves by harvesting kinetic energy from an activating environment \cite{Schweitzer,Reviews,Kroy}. Such a biomimetic counterpart of cellular motility is fueled by stationary non-equilibrium processes activated by the swimmers themselves through some built-in functional asymmetry \cite{Schweitzer,Reviews,Sen_rev}. Similarly to bacteria, artificial microswimmers also manifest tactic properties by diffusing up or down long-scale monotonic gradients in the activating suspension medium \cite{SenPRL,Sano,Baraban,materials,ChenChen}, thus exhibiting a positive or negative net tactic drift. As a major difference, bacteria control their response by means of complex finite-time adaptive mechanisms \cite{Armitage}, whereas synthetic swimmers, lacking
an internal structure \cite{rmpHM_2009,chemphyschem}, respond
instantaneously to the local activation properties of the surrounding medium, that is with no spatio-temporal memory \cite{Kapral}.

Controlling the transport of synthetic microswimmers in a spatio-temporally modulated activating medium is emerging as a key task in nanorobotics \cite{Sen_rev}. In a recent work \cite {ourPRE} we investigated the diffusive dynamics of microswimmers subjected to traveling activating pulses at low Reynolds numbers. We observed that small, point-like swimmers with positive taxis in a monotonic activation gradient actually drift towards or away from the pulse source, depending on the pulse speed and waveform. However, microswimmers of finite size immersed in an activation gradient can be subject to additive torques, which tend to align their self-propulsion velocities parallel or antiparallel to the gradient itself. The magnitude and sign of the ensuing (positive or negative) self-polarization effect have been shown to depend on the surface mobility of the nonuniform swimmer's coating \cite{Wurger,polar1}. In this paper we prove that tactic transport in a traveling activating wave can be selectively regulated by tuning the self-polarization properties of the chosen active swimmers. In particular, the magnitude of the tactic drift diminishes in the presence of positive self-polarization, to the point of causing tactic current reversals. Vice versa, negative self-polarization just enhances the tactic behavior reported in Ref.\ \citenum{ourPRE}.

\begin{figure*}[t]
\centering
\includegraphics[width=\textwidth*\real{0.8}*\real{\scalefac}]{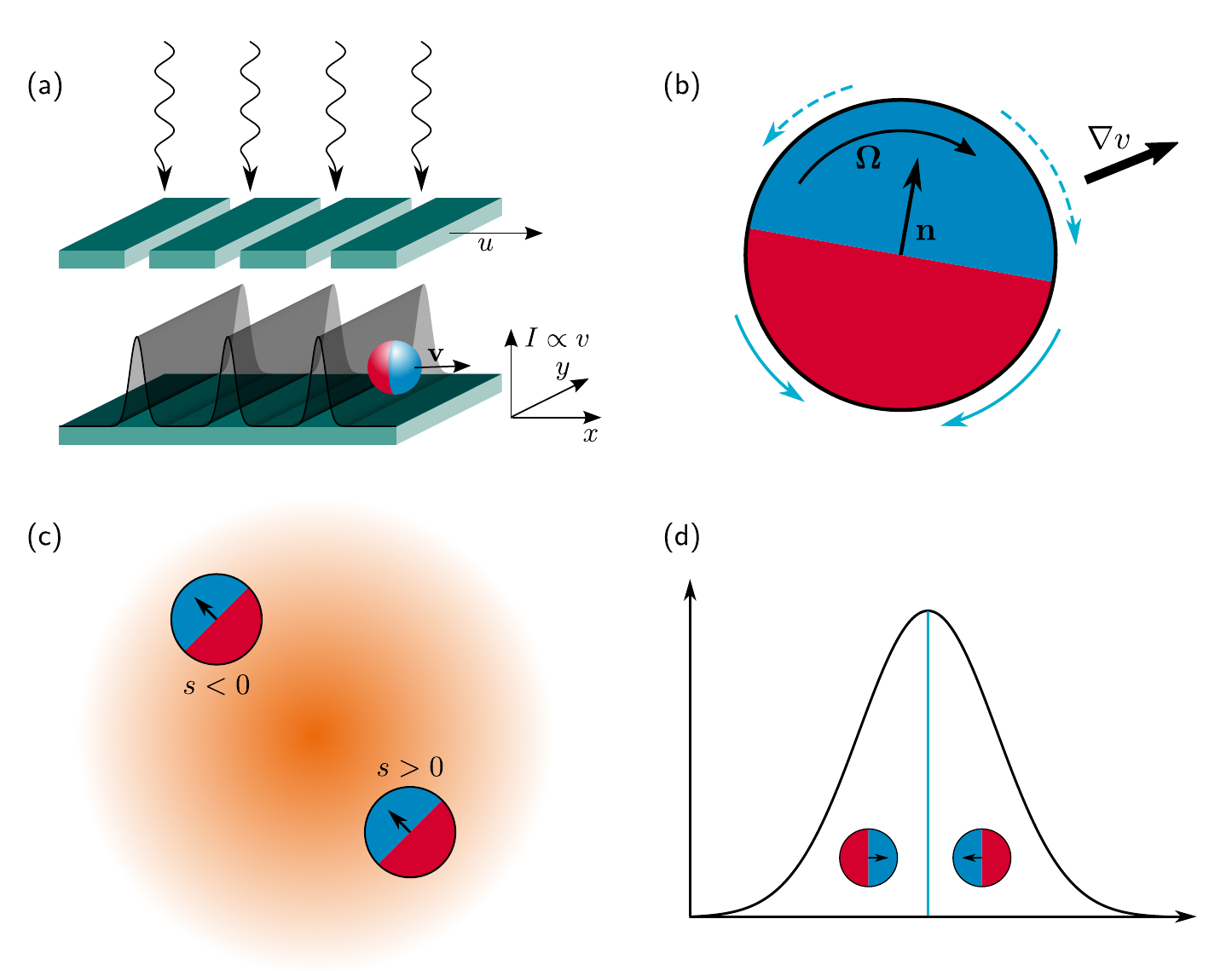}
\caption{ (a) Swimmer in a 2D traveling thermophoretic wave generated by a defocused laser beam of intensity $I$ \cite{Sano}. The activation modulation profile, $v(x)$, is tailored by inserting an appropriate slit screen between the light source and the substrate; the corresponding ADW is obtained by sliding the screen at a constant speed, $u$. (b) Self-polarization mechanism underlying Eq.\ \ref{Omega} with torque $\bm{\Omega}$ in an activation gradient $\bm{\nabla}v$. The different quasi-slip velocties on the coated (red) and uncoated (blue) face of a spherical Janus particle are denoted by solid and dashed curves arrows, respectively \cite{Wurger}. (c) Orientation of self-polarizing swimmers with $s<0$ and $s>0$ in a radially symmetric activating spot. (d) Trapping mechanism, where a positively self-polarizing swimmer with $s>0$ is sucked inside a traveling activation pulse. \label{F1}}
\end{figure*}

The paper is organized as follows. We initially summarize the tactic properties of artificial apolar microswimmers diffusing in a spatio-temporally modulated activation gradient \cite{ourPRE} (Sec.\ ``Taxis in active density waves''). In the following section we then introduce an additional ingredient, namely self-polarization \cite{Wurger} [cf. Fig.\ \ref{F1}(b)-(d)], which can influence swimmers' response to external gradients. In Sec.\ ``Taxis of self-polarizing microswimmers'' we investigate, both numerically and analytically, the interplay between such (possibly opposite) tactic mechanisms, the former controlled by rotational noise and the latter by self-polarization. In Figs. \ref{F2}-\ref{F5} we show how, under different parameter choices, one mechanism prevails upon the other, thus determining the sign of the resulting tactic drift. Finally, in Sec.\ ``Conclusions'' we make a few concluding remarks about future microfluidic devices.

\section*{Results and Discussion}

\subsection*{Taxis in active density waves}\label{chemo}

At low Reynolds numbers, the diffusion of an artificial {\it non-selfpolarizing}
microswimmer on a 2D substrate is modeled by a set of simple
Langevin equations (LE) \cite{Golestanian,Buttinoni,Bechinger,Nori},
\begin{eqnarray}
\dot x&=&v(x,t)\cos \phi +\sqrt{D_0}\; \xi_x(t),\nonumber \\ \dot
y&=&v(x,t)\sin
\phi+\sqrt{D_0}\; \xi_y(t), \nonumber \\
\dot \phi&=&\sqrt{D_\phi}\; \xi_\phi(t) \label{LE1},
\end{eqnarray}
with three fluctuating terms: two translational ones of intensity $D_0$, and an orientational one of intensity $D_\phi$. All noises are Gaussian and stationary, with zero mean and autocorrelation functions $\langle \xi_i(t)
\xi_j(0)\rangle =2\delta_{ij}\delta(t)$, where $i,j=x,y,\phi$. For the sake of generality, we treat $D_0$ and $D_\phi$ as independent parameters. The angle $\phi$ characterizes the swimmer's orientation and is measured with respect to the $x$ axis.

When the swimmer floats in a homogeneous activating suspension, its
propulsion speed is nearly constant, $v_0$. It then undergoes an active Brownian motion with a finite persistence time, $\tau_\phi=1/D_\phi$, length, $l_\phi=v_0\tau_\phi$, and bulk diffusion constant, $\lim_{t\to \infty}\langle
[x(t)-x(0)]^2\rangle/(2t)=D_0+D_s$, with $D_s=v_0^2/2D_\phi$ \cite{Golestanian}.

To model the effect of activating pulses sweeping through the
suspension fluid, we assume that the propulsion speed, $v(x,t)$, is a
local function of the activation properties of the medium at the swimmer's position\cite{Schnitzer}. Indeed, the propulsion speed of most artificial chemotactic swimmers grows linearly with the concentration of the chemoactivants in the
suspension, whereas, as assumed in Eqs.\ (\ref{LE1}), their angular
diffusion stays almost the same \cite{Sen_rev,Sano,Golestanian}. An example is represented by the idealized thermo-phoretic experiment sketched in Fig.\ \ref{F1}(a). Accordingly, pulses propagating with speed $u$ from left to right along the $x$ axis, are modeled by replacing $v(x,t)$ in Eqs.\ (\ref{LE1}) with an appropriate function $v(x,t)=v(x-ut)$ \cite{Schweitzer}.

\begin{figure*}[t]
\includegraphics[width=\textwidth*\real{0.84}*\real{\scalefac}]{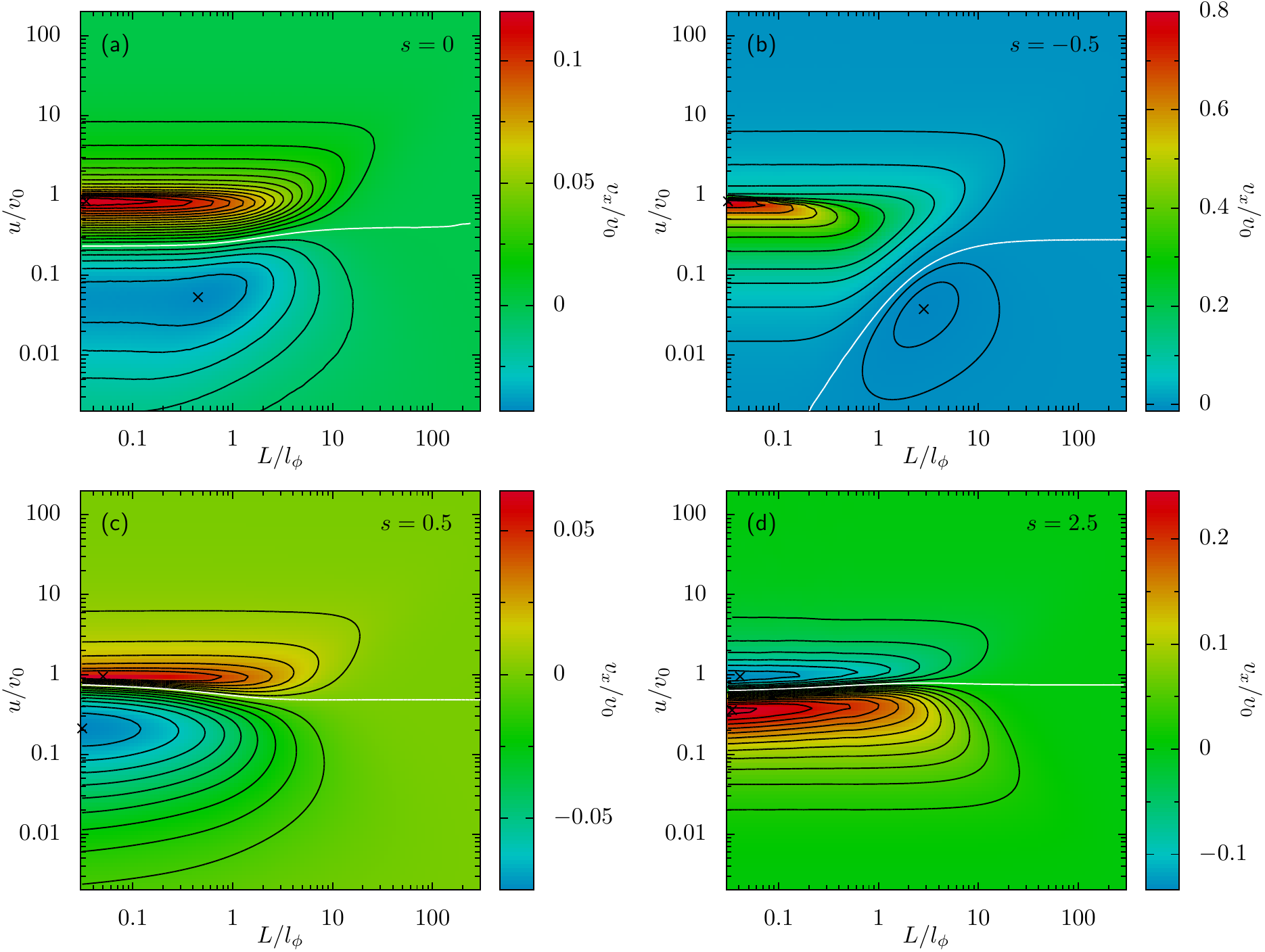}
\caption{Tactic response of a self-polarizing swimmer in the sinusoidal ADW (\ref{ADW}). The model parameters were chosen consistently
with experimental values of Ref.\ \citenum{Bechinger}: $v_0=53\,{\rm\mu m/s}$, and $D_\phi=165\,{\rm s^{-1}}$; only $D_0$ was set to zero for simplicity. All results were obtained by combining the numerical integration of the model LE's (\ref{LE1})---where the angular dynamics follow from Eq.\ (\ref{LE1-3})---and the numerical solution of the corresponding FPE (\ref{FPE}) \cite{ourPRE}. Contour plots of the longitudinal drift velocity, $v_x$, have been obtained for different self-polarization strengths, namely $s=0$ (a), $-0.5$ (b), $0.5$ (c) and $2.5$ (d).
\label{F2}}
\end{figure*}

For the sake of simplicity, we restrict our analysis to active density waves (ADW) with a spatially periodic waveform; i.e.,
\begin{equation} \label{ADW}
v(x, t=0)=v(x)= v_0 \sin^2(\pi x/L).
\end{equation}
Other activating pulse sequences have been addressed in Ref.\ \citenum{ourPRE}. Like in most experimental setups, we assume that the swimmer's propulsion parameters, $v_0$, and $\tau_\phi$ are fixed, whereas $D_0$ and the ADW parameters, $u$ and $L$, can be tuned at experimenter's convenience.

In the stationary regime, the swimmer's chemotaxis is characterized by the drift velocity,
$v_x=\langle \dot x \rangle=\lim_{t\to \infty}\langle
[x(t)-x(0)]\rangle/t$, plotted in Fig.\ \ref{F2}(a) for $D_0=0$ and in Fig.\ \ref{F5}(a) for $D_0>0$ and different values of the wave parameters. In the regime of zero translational noise, $D_0=0$, the regions of the plane
$(L,u)$ exhibiting positive or negative chemotaxis are delimited by a horizontal {\it separatrix} curve, located at around $u\sim v_0$. Negative chemotaxis is induced by slow ADWs, with $u<v_0$. Indeed, for fast ADWs, the distance a swimmer can travel
without crossing a minimum of $v(x)$ is longer to the right than to
the left. As a consequence of the fore-rear asymmetry of a traveling wave, one thus expects that $v_x>0$ (Stokes' drift \cite{stokes2}). Moreover, the distance the swimmer can surf a wave with $u \simeq v_0$ is limited solely by its persistence time, $\tau_\phi$, so that here positive chemotaxis is most pronounced for $l_\phi \gtrsim L$. Vice versa, for $u\ll v_0$ the swimmer diffuses under the wave crests until it hits one of the troughs, where $v(x)\to0$. Since the ADW propagates to the right, for $D_0=0$ the particle cannot cross a trough from left to right, hence $v_x<0$.

However, as discussed in Ref.\ \citenum{ourPRE}, translational fluctuations with $D_0>0$ allow the swimmer to diffuse across the ADW troughs in both directions, thus suppressing its negative tactic response for $u<v_0$. As a result, the separatrix between positive and negative taxis, almost horizontal in Fig.\ \ref{F2}(a), turns downward upon decreasing the ADW length, $L$, smaller than the swimmer's effective diffusive length \cite{ourPRE}, $D_0/v_0$. Further details on the role of translational noise on tactic transport are reported in Sec.\ ``Taxis of self-polarizing microswimmers'' for the more general case of self-polarizing swimmers.

\subsection*{Self-polarization in activation density gradients} \label{polar}

It should be noticed that the latter self-propulsion model disregards hydrodynamic effects. Such a simplification is motivated and corroborated by the following experimental circumstances, holding under homogeneous activation conditions: (i) the swimmers freely diffuse in the bulk away from the container's walls \cite{Ghosh,walls1,walls2,walls3}; (ii) their density can be lowered so as to avoid clustering \cite{Navarro}; and (iii) they can be fabricated rounded in shape and so small in size (i.e., almost pointlike) to minimize hydrodynamic backflow effects \cite{Uspal}. However, in the case of realistic asymmetric, finite-size swimmers, the activation gradients considered here are likely to produce substantial hydrodynamic effects in the form of an additional self-propulsion velocity term and, more importantly, a polarizing torque that tends to align the particle's velocity parallel or anti-parallel to the gradient \cite{Wurger}. We refer to the latter mechanism respectively as to positive or negative swimmer's self-polarization. The simulation results presented in the next section demonstrate that the tactic drift is enhanced for negative self-polarizing swimmers and suppressed in the opposite case, as also observed in Ref.\ \citenum{Sano}.

As shown in Ref.\ \citenum{Wurger}, an activation gradient influences both the magnitude and the orientation of the swimmer's velocity. In the latter section we assumed the swimmer's self-propulsion velocity to be proportional to the local active density. Correspondingly, in our notation the self-propulsion velocity ${\vec v}(x,t)$ would change by an amount $\delta {\vec v}= \gamma \bm{\nabla} v(x,t)$ and rotate due to the effective torque \cite{Wurger}
\begin{equation} \label{Omega}
\bm{\Omega}=s {\vec n}\times\bm{\nabla} v(x,t).
\end{equation}
Here, $\vec n$ is the particle's orientation vector and the parameters $\gamma$ and $s$ characterize the particle's surface coupling to the activation gradient. For the sake of simplicity, in the following we neglect the self-propulsion velocity correction, being typically $\delta v\ll v_0$ \cite{Wurger}, and focus our analysis on the polarization term of Eq.\ (\ref{Omega}). Accordingly, the third LE in Eq.\ (\ref{LE1}) incorporates the additional torque $\vert\bm{\Omega}\vert=\Omega(x,\phi)$, that is,
\begin{equation} \label{LE1-3}
\dot \phi= \Omega(x,\phi)+\sqrt{D_\phi}\; \xi_\phi(t).
\end{equation}
The dimensionless polarization strength, $s$, is predicted to depend on the geometry of the swimmer and the physio-chemical properties of its surface \cite{Wurger}. The torque $\bm{\Omega}$ vanishes for $\sin \phi=0$, causing the particle to orient itself either parallel ($\phi=0$) or antiparallel ($\phi=\pi$) to the gradient $\bm{\nabla} v(x,t)$. The stable orientation is positive, i.e., pointing inwards the pulse (negative, i.e., pointing outwards the pulse) for $s>0$ ($s<0$), as illustrated in Fig.\ \ref{F1}(c).

The transport properties of the model of Eqs.\ (\ref{LE1}) and (\ref{LE1-3}) have been investigated by numerically solving the relevant Fokker-Planck equation (FPE) \cite{ourPRE},
\begin{equation} \label{FPE}
\frac{\partial}{\partial t'}P=\left(\frac{D_\phi L}{v_0}\frac{\partial^2}{\partial \phi^2}+\frac{s}{v_0}\frac{dv(x')}{dx'}\frac{\partial}{\partial \phi}\sin \phi + \frac{D_0}{Lv_0}\frac{\partial^2}{\partial x'^2} -\cos \phi \frac{\partial}{\partial x'}\frac{v(x')}{v_0} + \frac{u}{v_0}\frac{\partial}{\partial x'} \right)P,
\end{equation}
for the reduced probability density function $P=P(x',\phi, t')$, obtained by introducing the dimensionless coordinates $x'=(x-ut)/L$ and $t'=v_0 t/L$ and integrating over the transverse coordinate $y$ \cite{ourPRE}.

For convenience and better readability, the prime sign will be dropped throughout the following, implying that $x$ and $t$ from hereon refer to the above-noted dimensionless coordinates. We remark here that the averaged drift velocity, $v_x$, reads, in terms of the {\it rescaled} particle's velocity in the co-moving frame as $v_x=v_0\langle \dot x\rangle + u$. The stationary solution, $P_{\rm st}(x,\phi)$, of the FPE (\ref{FPE}) obeys periodic boundary conditions both in $x$ and $\phi$, i.e., $P_{\rm st}(x+1,\phi+2\pi)=P_{\rm st}(x,\phi)$.

In preparation for the discussion of the influence of noise on the swimmer's tactic drift (see the following section), we notice that the natural units of $D_0$ and $D_\phi$ are, respectively, $Lv_0$ and $v_0/L$. This implies that upon increasing the ADW length, $L$, the effect of the rotational noise increases, while that of the translational noises is suppressed.

\subsection*{Taxis of self-polarizing microswimmers} \label{chemo-polar}

The qualitative effect of self-polarization on the swimmers' tactic response is summarized in Fig.\ \ref{F2}. As anticipated, negative self-polarization, $s<0$, enhances the magnitude of the positive tactic drift described in Sec.\ ``Taxis in active density waves'', without much affecting the sign of $v_x$. On the contrary, a positively self-polarized swimmer appears to invert the sign of its drift velocity as $s$ is raised across the threshold value $s=1$.

This behavior can be explained with the fact that in our model the fore and the rear of the ADW pulses induce opposite polarization responses.
In the case of positive self-polarization, $s>0$, the self-propelling particle gets oriented parallel to the $v(x)$ gradients, that is inward the travelling pulse, as illustrated in Fig.\ \ref{F1}(d). In the diffusive regime of the swimmer's dynamics, namely for comparatively slow ADW's with $u<v_0$, the effect of positive self-polarization can be so pronounced to dynamically trap the particle under a wave crest, see Fig.\ \ref{F1}(d). In conclusion, positive self-polarization induces an additional tactic response opposite to that of apolar swimmers as described above. A comparison of panels (a), (c) and (d) of Fig.\ \ref{F2}, clearly shows that the negative (positive) drift below (above) the separatrix gets suppressed on increasing $s$, panel (c), until it turns positive (negative), panel (d). Vice versa, on reversing the sign of $s$, panel (b), a self-polarizing particle is expelled faster from both sides of the traveling pulse. In the ballistic regime, $u>v_0$, this mechanism makes a swimmer with $s<0$ surf the advancing front of a traveling pulse for a longer time interval than its receding tail, so that the swimmer's tactic response grows more positive. By the same token, in the diffusive regime, $u<v_0$, the negative tactic drift of a negatively self-polarizing swimmer gets strongly suppressed, to the point that the separatrix slants downward.

The trapping mechanism of positively self-polarizing swimmers by a traveling ADW is clearly illustrated in Fig.\ \ref{F3} for the noiseless limit, $D_0=D_\phi=0$, and will be further discussed in the following.

\subsubsection*{The noiseless limit}\label{noiseless}

\begin{figure*}[t]
\includegraphics[width=\textwidth*\real{0.8}*\real{\scalefac}]{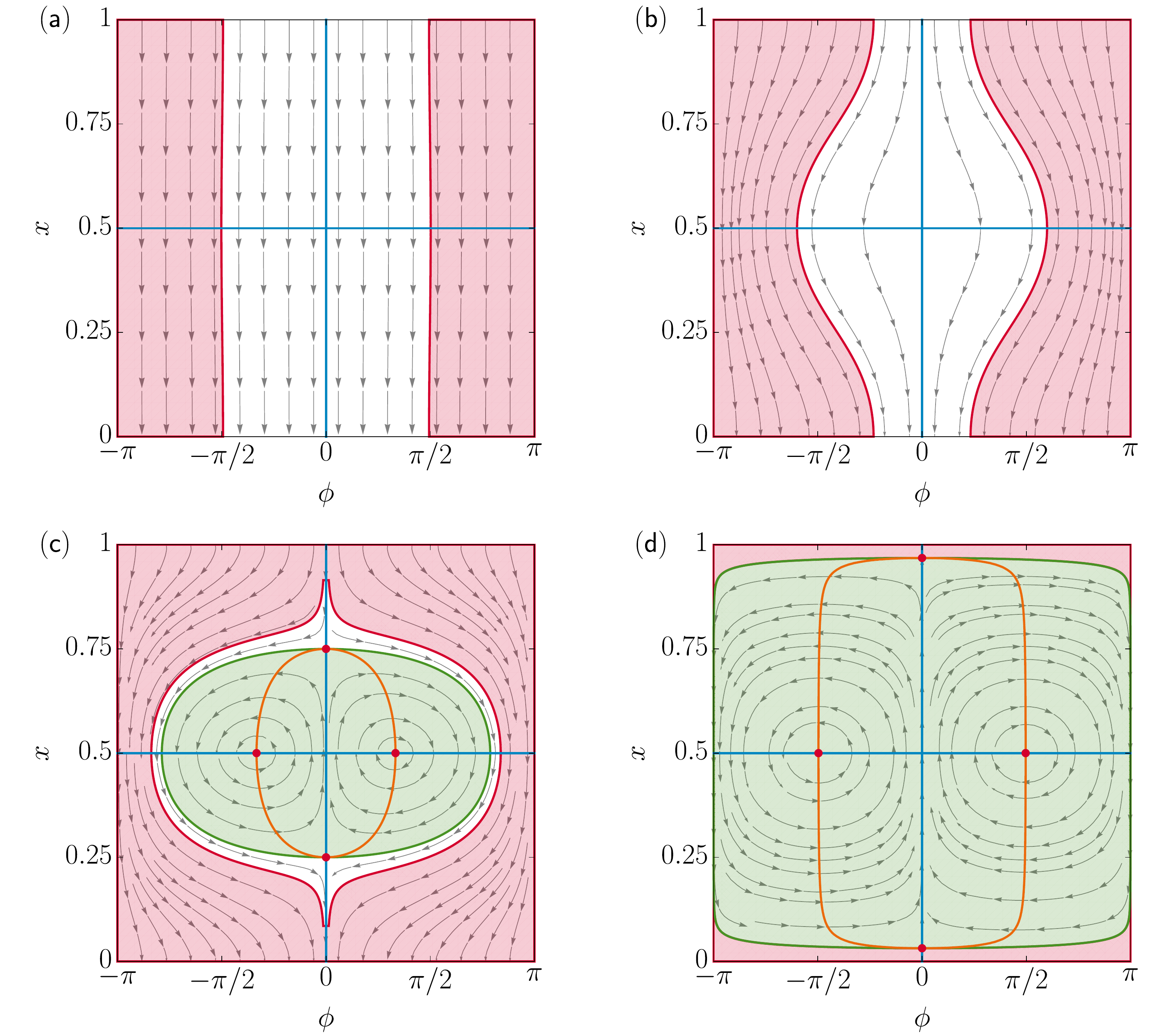}
\caption{Noiseless regime, $D_\phi=D_0=0$. Velocity field $(\dot x, \dot \phi)$, see Eq.\ (\ref{LEzero}), for $s = 2.5$ and $u/v_0 = 100$ (a), $2$ (b), $0.5$ (c), and $0.01$ (d). The orange and blue curves denote respectively the (existing) $x$ and $\phi$ nullclines of Eqs.\ (\ref{LEzero}); their intersections (red dots) locate two stable ($x=1/2$) and unstable ($\phi=0$) fixed points. Marked in red is the region of initial conditions corresponding to a negative drift velocity, i.e., $v_x=v_0\langle \dot x\rangle +u<0$, and in green the trapping region, i.e., $v_x=u$, where the particle oscillates inside the pulse in the co-moving frame (see text). For $s<0$ the velocity field is characterized by two symmetric saddle points with $x=1/2$, one attractor ($\phi=0$, $x>1/2$), and one repellor ($\phi=0$, $x<1/2$), not shown here. \label{F3}}
\end{figure*}

As mentioned above, positive self-polarization tends to suppress the tactic response of an apolar swimmer. In particular, for the sinusoidal ADW (\ref{ADW}), such competing effects cancel each other exactly at $s=1$. This can be readily proven by setting $s=1$ in the FPE (\ref{FPE}), which, after simple algebraic manipulations can thus be rewritten as
\begin{equation} \label{FPEs1}
\frac{\partial}{\partial t}P=\left(\frac{D_\phi L}{v_0}\frac{\partial^2}{\partial \phi^2}+\frac{\sin \phi}{v_0}\frac{dv(x)}{dx}\frac{\partial}{\partial \phi} + \frac{D_0}{Lv_0}\frac{\partial^2}{\partial x^2} -\cos \phi \frac{v(x)}{v_0}\frac{\partial}{\partial x} + \frac{u}{v_0}\frac{\partial}{\partial x} \right)P.
\end{equation}
On imposing the periodic boundary conditions detailed at the bottom of Sec.\ ``Self-polarization in activation density gradients'', one obtains immediately $P_{\rm st}(x,\phi)=1/(2\pi)$, so that the corresponding tactic drift must vanish, i.e., $v_x=0$ for any choice of the ADW parameters $L$ and $u$. It is only for $s>1$ that the drift velocity switches sign in the entire $(L,u)$ plane, as illustrated in Fig.\ \ref{F2}.

To clarify the interplay between the tactic mechanism of Sec.\ ``Taxis in active density waves'' and self-polarization, we address now in more detail the purely deterministic case with $D_0=D_\phi=0$. We restrict our analysis to the more interesting case of positive self-polarization, $s>0$, where the tactic drift reversal occurs. A generalization of our approach to self-polarizing swimmers with $s<0$ is straightforward. For the sinusoidal ADW of Eq.\ (\ref{ADW}), the set of LE's (\ref{LE1}) turns into a set of nonlinear ordinary differential equations, namely,
\begin{alignat}{2}
\dot{\phi}&=-s\pi\sin(\phi)\sin(2\pi x)\nonumber\\
\dot{x}&=\cos(\phi)\sin^2(\pi x)-u/v_0.
\label{LEzero}
\end{alignat}

The drift velocity of a strongly self-polarizing deterministic swimmer is plotted in Fig.\ \ref{F4}(a). The $u$-dependence of $v_x$ can be explained as follows. In the co-moving wave-frame with $u\to\infty$, the particle is pulled through the activating pulses so fast that its dynamics is insensitive to the modulation of the activation gradient. In such a limit, the velocity field, $(\dot x, \dot \phi)$, consists of parallel lines with $\dot x=-u/v_0$ and $\dot \phi=0$, see Fig.\ \ref{F3}(a); hence $v_x\to 0$.

Upon lowering the ADW speed, $u$, the swimmer begins to perceive the modulated gradients, as it crosses the wave pulses. Accordingly, the velocity field gets progressively distorted as shown in panels (b)-(d) of Fig.\ \ref{F3}. As long as $u>v_0$, Fig.\ \ref{F3}(b), the field lines are no longer parallel, but still cross the reduced ADW unit cell $(0,1)$ from top to bottom with $\dot x<0$, while the majority of field lines starting at $x=1$ crosses the centerline, $x=1/2$, with $|\phi|>\pi/2$. This means that a positively self-polarizing swimmer points preferably to the left ($\Rightarrow\dot x<-u/v_0$) under the ADW crests and to the right ($\Rightarrow\dot x>-u/v_0$) in the troughs.
Since the regions where the particle preferably points to the left (right) correspond to high (low) self-propulsive velocities, on switching back to the laboratory frame, for $u>v_0$ the swimmer's net drift is expected to be negative, in agreement with the data reported in Fig.\ \ref{F4}(a).

For ADW speeds smaller than the swimmer's self-propulsion speed, $u<v_0$, the swimmer gets trapped by the traveling pulses with either positive or negative orientation. Such a mechanism is marked by the emergence of two periodic bound solutions of Eqs.\ (\ref{LEzero}) with $s>0$, see Figs.\ \ref{F3}(c),(d). Closed swimmer's trajectories wind up around the stable fixed points $(x_\mathrm{f},\phi_\mathrm{f})=(1/2,\pm\arccos(u/v_0))$ with frequencies $\pm\pi\sqrt{2s\left(1-(u/v_0)^2\right)}$, as one can easily prove by rewriting the coupled Eqs.\ (\ref{LEzero}) in linear approximation,
\begin{equation}
\frac{d}{dt}\begin{pmatrix}\tilde{\phi}\\\tilde{x}\end{pmatrix}=
\pm\sqrt{1-\left(\frac{u}{v_0}\right)^2}\begin{pmatrix}\phantom{-}0&2\pi^2s\\-1&0\end{pmatrix}\begin{pmatrix}\tilde{\phi}\\\tilde{x}\end{pmatrix},
\end{equation}
with $\tilde{\phi}=\phi-\phi_\mathrm{f}$ and $\tilde{x}=x-x_\mathrm{f}$.

Of course, initial conditions in the region of the configuration space $(x, \phi)$ corresponding to such closed trajectories, contribute the maximum positive amount, $u$, to the net drift velocity of the noiseless swimmer. Moreover, as illustrated in Fig.\ \ref{F3}(d), upon lowering $u$, the trapping region becomes larger and larger, until $v_x$ finally turns positive. Accordingly, one predicts that the overall drift velocity $v_x$ must tend to zero proportional to $u$ in the limit $u\to 0$, see Fig.\ \ref{F4}(a).

\subsubsection*{The role of rotational noise}\label{Dphi}

We now consider the more realistic case when the rotational noise, $\xi_\phi(t)$ in Eq.\ (\ref{LE1-3}), has finite intensity, $D_\phi>0$, and the translational noises of the LE's (\ref{LE1}) are comparatively so weak to be safely neglected, i.e., $D_0=0$. The swimmer then tends to rotate as a combined effect of the polarization torque $\bm{\Omega}$ of Eq.\ (\ref{Omega}) and the angular fluctuations due to thermal Brownian motion in the suspension fluid and, possibly, the particle's self-propulsion mechanism itself. In dimensionless units, the time constants for these two orientational processes can be readily estimated, respectively, from Eq.\ (\ref{LEzero}), $\tau^{-1}_s=\pi s$, and Eq.\ (\ref{LE1}), $\tau^{-1}_\phi=LD_\phi/v_0$.

\begin{figure}[t]
\includegraphics[width=\textwidth*\real{0.95}*\real{\scalefac}]{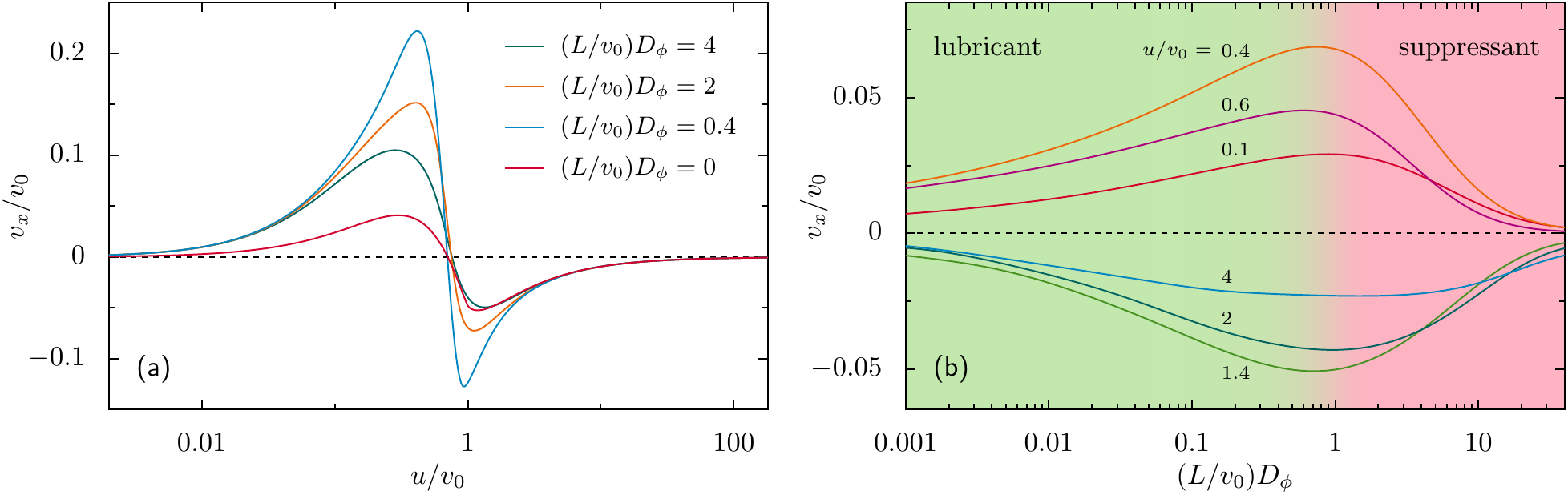}
 \caption{Role of the rotational noise. Panel (a): Drift velocity of the strongly self-polarizing swimmer with $s=2.5$, Fig.\ \ref{F2}(d), versus $u$, both in units of $v_0$, for different values of $(L/v_0)D_\phi$ and $D_0=0$, Panel (b): Drift velocity of the strongly self-polarizing swimmer of Fig.\ \ref{F5}(d) versus $D_\phi$, both in the dimensionless units, for $v_0=53 {\rm \mu m/s}$, $D_0=2.2\,{\rm \mu m^2/s}$, and different $u$. The data plotted here have been obtained by numerically integrating the LE's (\ref{LE1}) and (\ref{LE1-3}) or the FPE (\ref{FPE}). \label{F4}}
\end{figure}

In Fig.\ \ref{F4}(b) we plotted the tactic drift velocity of the strongly positively self-polarizing swimmer of Fig.\ \ref{F2}(d) versus the intensity of its rotational noise (which, in dimensionless units, coincides with $\tau^{-1}_\phi$). A small amount of translational noise was added to simulate a more realistic physical situation, but plays no role in the present discussion. As argued above, in the noiseless limit, $v_x$ is positive for $u<v_0$ and negative for $u>v_0$. On raising the level of rotational noise, we notice that the modulus of $v_x$ first increases until it reaches a maximum; above such an optimal $D_\phi$ value, $|v_x|$ starts decreasing until it vanishes altogether. Moreover, we observed that adding any amount of rotational noise appears not to change the sign of $v_x$. The drift velocity thus exhibits a prominent resonant behavior for both positive and negative tactic responses.

Such a remarkable noise dependence has a certain similarity with the phenomenology of stochastic resonance \cite{RMP98}. A small amount of rotational noise acts as a {\it lubricant}, meaning that it favors the swimmer's reorientation with respect to the activation gradients. In particular, for a positively self-polarizing swimmer, this anticipates the onset of the trapping mechanism described above in the noiseless limit. Note that as long as the tactic mechanisms controlled by self-polarization and rotational noise can be regarded as independent, the effective rotational time constant is of the order of $\tau=\tau_s \tau_\phi/(\tau_s+\tau_\phi)$, with $\tau \to \tau_s$ for $D_\phi \to 0$ and $\tau \to \tau_\phi$ for $D_\phi \to \infty$.

On the contrary, large angular fluctuations tend to randomize the swimmer's trajectories, so that the regular patterns of field lines displayed in Fig.\ \ref{F3} get blurred until they eventually become undiscernible. Under these circumstances, noise acts rather as a {\it suppressant} of the tactic response and, accordingly, the modulus of the swimmer's drift velocity vanishes.

\subsubsection*{The role of translational noise}\label{Dzero}

\begin{figure*}[t]
\includegraphics[width=\textwidth*\real{0.84}*\real{\scalefac}]{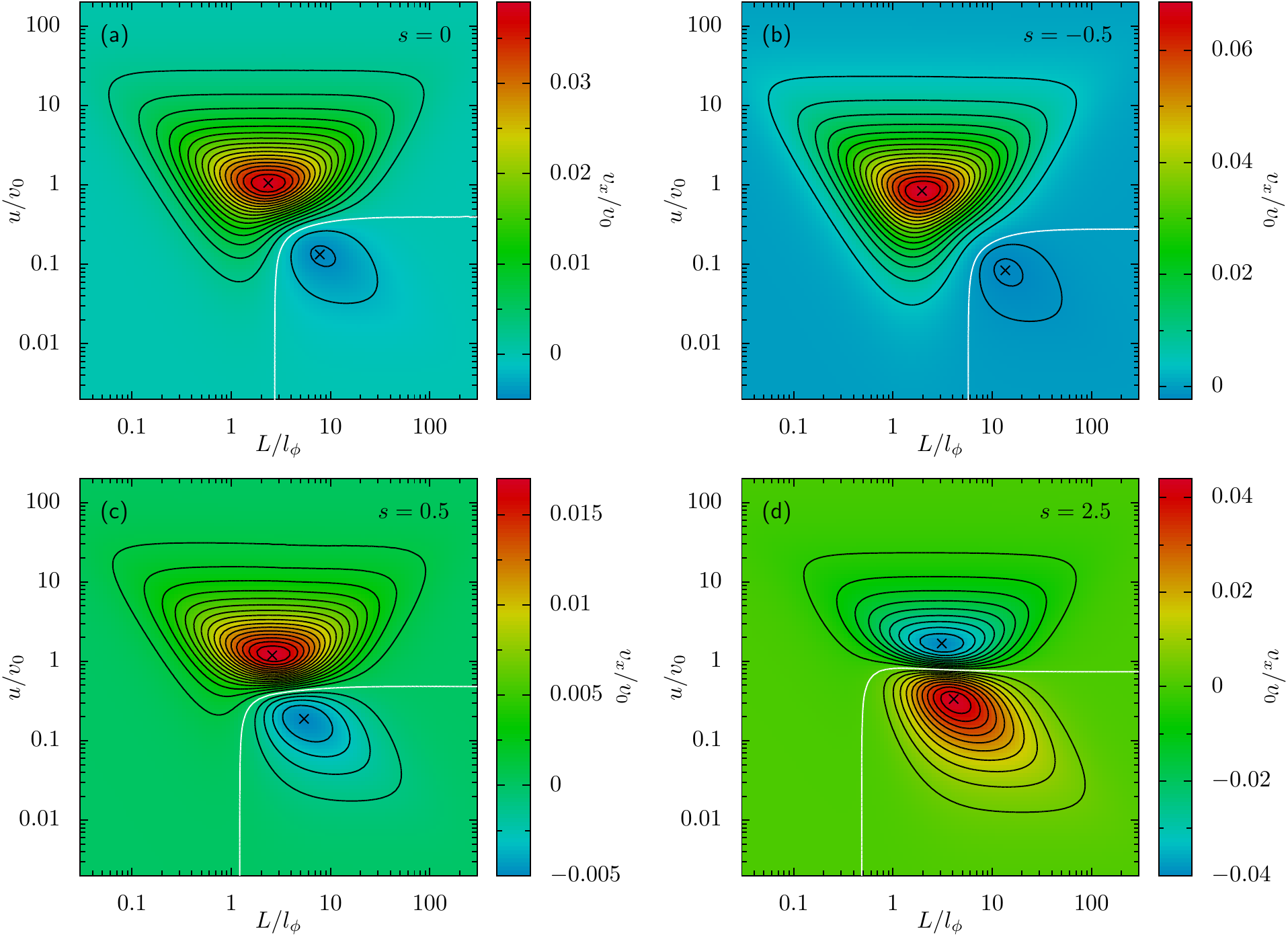}
 \caption{Role of the translational noise. Contour plots of the longitudinal drift velocity $v_x$ for the same model parameters as in Fig.\ \ref{F2}, but for finite translational noise, $D_0=2.2\,{\rm \mu m^2/s}$. \label{F5}}
\end{figure*}

In Sec.\ ``Taxis in active density waves'' we anticipated that for apolar swimmers, $s=0$, the noise sources in the first two LE's (\ref{LE1}) impact the curve separating the $(u,L)$ regions with positive and negative tactic drift. In fact,
the translational fluctuations help the swimmer to diffuse across the
ADW troughs. Accordingly, with increasing $D_0$ the negative
rectification effect gets suppressed with respect to the positive Stokes' drift, until eventually $v_x$ changes sign. Of course, the same effect can
also be achieved by shortening the ADW length, $L$, see the remarks following Eq.\ (\ref{FPE}). This is why in Fig.\ \ref{F5}(a) the separatrix curve bends downward, almost vertically, at a critical value of $L$. The dependence of the vertical branch of the separatrix on the model parameters was analyzed in Ref.\ \citenum{ourPRE}. In particular, we showed that its position along the horizontal axis shifts to the right proportional to $D_0$ and to the left proportional to a possible offset of the wave pulse ($w_0$ in Ref.\ \citenum{ourPRE}), independently of $l_\phi$. Such a behavior hints at the existence of a critical value of the dimensionless ratio $D_0/Lv_0$, above which negative taxis is suppressed \cite{ourPRE}.

We consider now the more general case of self-polarizing swimmers, $s\neq 0$. On comparing Figs.\ \ref{F2} and \ref{F5}, we observe that in Fig.\ \ref{F5} the bottom-left quadrants of all contour plots with $s<1$ display positive tactic velocities. For $s>1$, instead, the sign of the tactic velocity gets reversed, reminiscent of the tactic reversal discussed in the noiseless limit; accordingly, for $D_0>0$, the positive $v_x$ values are then restricted to the bottom-right quadrants of the contour plots [the case for $s=2.5$ is shown in Fig.\ \ref{F5}(d)]. Thus, for $s>1$ the swimmer's tactic drift is indeed stronger influenced by self-polarization than by rotational noise, but translational fluctuations tend to still suppress it. The relevant separatrix curves depend on the polarization strength; in particular, their vertical branches are observed to shift to the left with increasing $s$ throughout the range we explored.


\subsection*{Conclusions} \label{conclusions}

As discussed in Ref.\ \citenum{ourPRE}, taxis of artificial microswimmers is a generally robust phenomenon. Particularly, laboratory demonstrations are already available \cite{SenPRL,Sano,Baraban,ChenChen,Armitage} and applications to
nanotechnology and medical sciences appear promising \cite{Sen_rev}. The tactic response of an artificial microswimmer results from its ability to diffuse in a spatio-temporally modulated activating medium, the direction and magnitude of the induced drift strongly depending on both the deterministic and random properties of the particular self-propulsion mechanism at work. In this report we have focused on the self-polarization of self-propelled particles, which we demonstrated to impact their ability to reorient themselves in response to local activation gradients. In a recent paper \cite{BechLoew}, appeared after the completion of the present work, a compelling experimental observation of the self-polarization effects anticipated here is reported for microswimmers with $s<0$.

With future work we aim to study the case of self-polarizing swimmers confined to narrow channels with corrugated walls, as the next step toward the design and operation of microfluidic devices powered by spatio-temporally modulated activation densities. We expect that the hydrodynamic interactions of the active swimmers with the walls likely will provide an additional and possibly competing orientational mechanism \cite{Uspal,tongji}, which contributes to regulating the swimmers' flow in a microfluidic circuit.

\bibliography{sample}

\section*{Acknowledgements}

This work has been supported by the cluster of excellence Nanosystems Initiative Munich (PH). PH and
FM acknowledge a financial support from the Center for Innovative
Technology (ACIT) of the University of Augsburg. FM also thanks the
DAAD for a Visiting Professor grant.

\section*{Author contributions statement}

A.G. performed all numerical calculations in this project. All authors contributed to the planning and the
discussion of the results as well as to the writing of this work.

\section*{Additional information}

Competing financial interests: The authors declare no competing financial interests.

\end{document}